\documentclass[11pt,english,aps,preprint,nofootinbib]{revtex4}
\usepackage[T1]{fontenc}
\usepackage[latin9]{inputenc}
\usepackage{amsmath}
\usepackage{amssymb}
\usepackage{esint}
\usepackage{babel}
\usepackage{comment}
\usepackage{color,graphicx}

\begin{document}
\title{Bohmian Mechanics, Collapse Models and the emergence of Classicality}

\author{Marko Toro\v{s}}

\email{marko.toros@ts.infn.it}

\affiliation{Department of Physics, University of Trieste, 34151 Miramare-Trieste,
Italy Istituto Nazionale di Fisica Nucleare, Sezione di Trieste, Via
Valerio 2, 34127 Trieste, Italy}

\author{Sandro Donadi}

\email{sandro.donadi@ts.infn.it}

\affiliation{Department of Physics, University of Trieste, 34151 Miramare-Trieste,
Italy Istituto Nazionale di Fisica Nucleare, Sezione di Trieste, Via
Valerio 2, 34127 Trieste, Italy}

\author{Angelo Bassi}

\email{bassi@ts.infn.it}

\affiliation{Department of Physics, University of Trieste, 34151 Miramare-Trieste,
Italy Istituto Nazionale di Fisica Nucleare, Sezione di Trieste, Via
Valerio 2, 34127 Trieste, Italy}
\begin{abstract}
We discuss the emergence of classical trajectories in Bohmian Mechanics (BM), when a macroscopic object interacts with an external environment. We show that in such a case the conditional wave function of the system follows a dynamics which, under reasonable assumptions, corresponds to that of the Ghirardi-Rimini-Weber (GRW) collapse model. As a consequence, Bohmian trajectories evolve classically. Our analysis also shows how the GRW (istantaneous) collapse process can be derived by an underlying continuous interaction of a quantum system with an external agent, thus throwing a light on how collapses can emerge from a deeper level theory.

\end{abstract}
\maketitle

\section{Introduction}
The superposition principle lies at the very heart of Quantum Mechanics.
This leads, as it was recognized very early, to paradoxical descriptions
for macroscopic systems, like the existence of macroscopic
linear superposition for arbitrarily large objects~\cite{1935NW.....23..807S}. This issue has given
rise to many different interpretations and some alternative theories
of Quantum Mechanics. In this article we focus on two of them, Bohmian
Mechanics and Collapse models, and we highlight new connections among
the two. 

Bohmian mechanics gives a rigorous mathematical description together
with a clear and unambiguous picture of the motion of quantum objects, in the
non relativistic regime, by supplementing
the wave function with the motion of actual particles' positions~\cite{durr}. For microscopic objects it reproduces the predictions of ordinary
quantum mechanics, while for macroscopic objects it is argued that
it reduces to classical mechanics~\cite{1464-4266-4-4-344,app}.
One of the main goal of this article is to investigate this claim more
in detail.

Collapse models modify the Schr\"odinger equation by adding stochastic
and nonlinear terms. This modification is negligible for microscopic
objects, while for macroscopic objects it forces the center of mass wave
function to become well localized in space. In this way one re-obtains,
for all practical purpose, the predictions of ordinary Quantum Mechanics
for microscopic systems, while macroscopic objects are effectively
described by almost point-like wave functions in space. As for Bohmian
Mechanics, the measurement problem is solved. 

The similarities of Bohmian Mechanics and Collapse models have already
been investigated in the literature~\cite{Allori01092008}. In particular, the hidden variable interpretation of collapse models was discussed in~\cite{1751-8121-44-27-275303,1751-8121-44-47-478001,laloe}. In this paper, we will unravel a much deeper connection among the two theories, and will present the following two results. 

%Bohmian mechanics, unlike ordinary Quantum Mechanics, provides a rigorous
%notion for the wave function of a sub-system of a given system i.e. the conditional wave
%function. We show a general relation between the conditional wave
%function and the reduced density matrix of a sub-system (Section \ref{sec:Bohmian-Mechanics}).

We introduce a new stochastic equation for the Ghirardi-Rimini-Weber (GRW) collapse model. This stochastic equation, from a mathematical point of view, is much simpler than the one recently proposed in the literature~\cite{PhysRevA.90.062135} and from a physical point of view it is closer to the original idea of GRW and J.S. Bell (Section \ref{sec:Collapse-models}).

More importantly, we discuss the classical limit of Bohmian Mechanics. First, we show how, when a single particle interacts with an external bath, its conditional wave function collapses. Second, we argue that this dynamics, in a well defined limit (istantaneous interaction with bath particles), is equivalent to the GRW dynamics. Third, we show that for a macroscopic many-body object, the center of mass wave function collapses according to the GRW dynamics with an amplified collapse rate. We finally argue how the GRW dynamics, when the object involved is large enough, induces a dynamics for the conditional wave function such that the Bohmian trajectories evolve classically (Sections \ref{sec:The-classical-limit} and \ref{classicallimit2}).

\section{Bohmian Mechanics\label{sec:Bohmian-Mechanics}}
In Bohmian Mechanics~\cite{1927JPhRa...8..225D,PhysRev.89.458,PhysRev.85.166,bohmbook,dgzbook,dgzarticle}, particles are described by points $\mathbf{X}_{1}(t),...,\mathbf{X}_{N}(t)$ moving in space under the influence of the wave function:  
\begin{equation}
\psi(\mathbf{x}_{1},...,\mathbf{x}_{N},t).
\end{equation}
Notice the difference between the actual positions of the particles (always denoted by capital letters) and the dependence of the wave function on the points of configuration space (always denoted by small letters). 
The wave function evolves according to the usual Schr\"odinger equation:
\begin{equation}
i\hbar\frac{\partial\psi(\mathbf{x}_{1},...,\mathbf{x}_{n},t)}{\partial t}=\sum_{j=1}^{N}-\frac{\hbar^{2}}{2m_{j}}\boldsymbol{\nabla}_{j}^{2}\psi(\mathbf{x}_{1},...,\mathbf{x}_{N},t)+\hat{V}\psi(\mathbf{x}_{1},...,\mathbf{x}_{N},t),\label{eq:Schrodinger}
\end{equation}
while the particles' positions evolve according to the guidance equation:
\begin{equation}
\frac{d\mathbf{X}_{j}(t)}{dt}=\mathbf{v}_{j}^{\psi}(\mathbf{X}_{1},...,\mathbf{X}_{N},t),\label{eq:velocity}
\end{equation}
where the form of velocity field is determined by imposing that the continuity equation
(quantum flux equation):
\begin{equation}
\partial_t|\psi|^{2} +\sum_{j=1}^{N}\boldsymbol{\nabla}_j\cdot(\mathbf{v}_{j}^{\psi}|\psi|^{2})=0\label{eq:continuity}
\end{equation}
must hold true. 

If the interaction potential $\hat{V}$ depends only on the position operators $\hat{\mathbf{x}}_{j}$ of the particles, then the velocity field
is given by:
\begin{equation}\label{velvx}
\mathbf{v}_{j}^{\psi}=\frac{\hbar}{m_{j}}\textrm{Im}\frac{\boldsymbol{\nabla}_j \psi}{\psi}.
\end{equation}
In general, as we will see, Eq.~(\ref{velvx}) does not hold if the
potential depends also on the momentum operators $\hat{\mathbf{p}}_{j}$.
In such a case, the velocity field has to be derived directly from Eq.~(\ref{eq:continuity}).

Besides determining the particles' motion, the wave function plays another fundamental role in Bohmian
mechanics. Given an ensemble of identically prepared systems, and assuming that the particles are distributed according to $|\psi|^2$ at an initial time (Quantum Equilibrium Hypothesis):
\begin{equation}
\mathbb{P}(\mathbf{X}_{1}(0)=\mathbf{x}_{1},...,\mathbf{X}_{N}(0)=\mathbf{x}_{N})=|\psi(\mathbf{x}_{1},...,\mathbf{x}_{N},0)|^{2},\label{eq:qeh}
\end{equation}
then, as a consequence of Eq.~(\ref{eq:continuity}):
\begin{equation}\label{eq:qeh-1}
\mathbb{P}(\mathbf{X}_{1}(t)=\mathbf{x}_{1},...,\mathbf{X}_{N}(t)=\mathbf{x}_{N})=|\psi(\mathbf{x}_{1},...,\mathbf{x}_{N},t)|^{2},\;\;\;\;\;\forall\,t>0.
\end{equation}
This makes sure that, for any time $t$, the predictions of Bohmian
Mechanics agree with those of ordinary Quantum Mechanics. 
%(when there are no ambiguities)

Let us consider $N + M$ particles described by the total
wave function $\psi(\mathbf{x}_{1},...,\mathbf{x}_{N},\mathbf{y}_{1},...,\mathbf{y}_{M},t)$ and
particle positions $\mathbf{X}_{1}(t),...,\mathbf{X}_{N}(t),\mathbf{Y}_{1}(t),...,\mathbf{Y}_{M}(t)$. What we have in mind is a sub-system (e.g. a rigid body) composed of $N$ particles, which interacts with an environment made of $M$ particles. In order to have a more concise and
readable notation, we introduce the vector variables $\{\mathbf{x}\}:=(\mathbf{x}_{1},...,\mathbf{x}_{N})$,
$\{\mathbf{y}\}:=(\mathbf{y}_{1},...,\mathbf{y}_{M})$ and the vector positions $\{\mathbf{X}(t)\}:=(\mathbf{X}_{1}(t),...,\mathbf{X}_{N}(t))$,
$\{\mathbf{Y}(t)\}:=(\mathbf{Y}_{1}(t),...,\mathbf{Y}_{M}(t))$.

The conditional wave function of the sub-system is defined as~\cite{dgzarticle}:
\begin{equation}
\psi_{\textrm{C}}(\{\mathbf{x}\},t)=\frac{\psi(\{\mathbf{x}\},\{\mathbf{Y}(t)\},t)}{||\psi(\{\mathbf{x}\},\{\mathbf{Y}(t)\},t)||_{\textrm{S}}},\label{eq:conditional}
\end{equation}
where 
\begin{equation}
||\psi(\{\mathbf{x}\},\{\mathbf{Y}(t)\},t)||_{\textrm{S}}=\int d\{\mathbf{x}\}|\psi(\{\mathbf{x}\},\{\mathbf{Y}(t)\},t)|^{2}\label{eq:bohmian_pdf_subsystem}
\end{equation}
and where $\int d\{\mathbf{x}\}:=\int d\mathbf{x}_1\int d\mathbf{x}_2...\int d\mathbf{x}_N $. We see that the variables $\{\mathbf{y}\}$ are replaced by the actual positions
of the environmental particles $\{\mathbf{Y}(t)\}$ and the resulting
wave function is normalized.

Given the density matrix in the position representation
\begin{equation}
\rho(\{\mathbf{x}\},\{\mathbf{x}'\};\{\mathbf{y}\},\{\mathbf{y}'\},t)=\psi(\{\mathbf{x}\},\{\mathbf{y}\},t)\psi(\{\mathbf{x}'\},\{\mathbf{y}'\},t)^{*},
\end{equation}
the reduced density matrix of the sub-system is defined as:
\begin{equation}
\rho_{\textrm{S}}(\{\mathbf{x}\};\{\mathbf{x}'\},t)=\textrm{Tr}_{\textrm{E}}\left[\rho(\{\mathbf{x}\},\{\mathbf{x}'\};\{\mathbf{y}\},\{\mathbf{y}'\},t)\right]:=\int d\{\mathbf{y}\}\,\rho(\{\mathbf{x}\},\{\mathbf{x}'\};\{\mathbf{y}\},\{\mathbf{y}\},t).\label{eq:reduced}
\end{equation}

There exists a simple and natural relation between the conditional wave function $\psi_{\textrm{C}}$
of the sub-system and its reduced density matrix $\rho_{\textrm{S}}$~\cite{densitymatrices}. From the definition of the conditional wave function we first define the conditional density matrix:
\begin{equation}
\rho_{\textrm{C}}(\{\mathbf{x}\};\{\mathbf{x}'\},t):=\psi_{\textrm{C}}(\{\mathbf{x}\},t)\psi_{\textrm{C}}(\{\mathbf{x}'\},t)^{*}
\end{equation}
We then consider an ensemble of identically prepared systems, whose particle positions are distributed according to Eq.~(\ref{eq:qeh-1}). Since the conditional wave function for the sub-system S depends on the actual positions $\{\mathbf{Y}\}$ of the environmental particles, the natural way to associate a density matrix to this ensemble is:
\begin{equation}\label{eq:theorem}
\mathbb{E}_{\textrm{E}}[\rho_{\textrm{C}}(\{\mathbf{x}\};\{\mathbf{x}'\},t)]
=\int d\{\mathbf{y}\}\, \mathbb{P}(\{\mathbf{Y}(t)\}=\{\mathbf{y}\})\frac{\psi(\{\mathbf{x}\},\{\mathbf{y}\},t)\psi(\{\mathbf{x}'\},\{\mathbf{y}\},t)^*}{||\psi(\{\mathbf{x}\},\{\mathbf{y}\},t)||_{\textrm{S}}^{2}},
\end{equation}
where $\mathbb{P}(\{\mathbf{Y}(t)\}=\{\mathbf{y}\})=\mathbb{P}(\mathbf{Y}_{1}(t)=\mathbf{y}_{1},...,\mathbf{Y}_{M}(t)=\mathbf{y}_{M})$
is the joint probability density for the environmental particles' positions. According
to Eq.~(\ref{eq:qeh-1}) we have $\mathbb{P}(\mathbf{Y}_{1}(t)=\mathbf{y}_{1},...,\mathbf{Y}_{M}(t)=\mathbf{y}_{M})=\int d\mathbf{x}_{1}...d\mathbf{x}_{N}|\psi(\mathbf{x}_{1},...,\mathbf{x}_{N},\mathbf{y}_{1},...,\mathbf{y}_{M},t)|^{2}=||\psi(\{\mathbf{x}\},\{\mathbf{y}\},t)||_{s}^{2}$, implying:
\begin{equation}
\rho_{\textrm{S}}(\{\mathbf{x}\};\{\mathbf{x}'\},t)=\mathbb{E}_{\textrm{E}}[\rho_{\textrm{C}}(\{\mathbf{x}\};\{\mathbf{x}'\},t)].
\end{equation}
Therefore, given a sub-system S, the bohmian conditional density matrix averaged over the environment is exactly the reduced density matrix in standard Quantum Mechanics, obtained from the total density matrix after tracing over the degrees of freedom
of the environment.

\section{Collapse models\label{sec:Collapse-models}}

We limit the presentation to the GRW model~\cite{PhysRevD.34.470}. For the other collapse models, one can refer to~\cite{Bassi2003257,RevModPhys.85.471}. In particular, we present the formulation of the GRW model due to J.S. Bell~\cite{bell}. We start by introducing a probability space $(\Omega,\mathcal{F},\mathbb{Q})$, where a $\mathbb{Q}$-Poisson process $N_{t}$ is defined. The Poisson process has mean
\begin{equation}
E_{\mathbb{Q}}[N(t)]=\lambda t,
\end{equation}
where $\lambda$ will set the strength of the collapse mechanism. The
standard value for $\lambda$ suggested by Ghirardi, Rimini and Weber
is $\lambda=10^{-16}\;\text{{s}}^{-1}$. It can be shown that the collapse rate, for the center of mass wave
function of a $N$-particle system, rescales to $\Lambda=N\lambda$. This is a manifestation of the so called ``amplification mechanism". 

The GRW dynamics assumes that the wave function collapses, at random times and in random points of the space. In between two collapses, the evolution is given by the Schr\"odinger equation. More formally, the GRW model assumes that during a collapse the state of a single particle changes instantaneously according to the following prescription:
\begin{equation}\label{collapsegrw}
\left|\psi(t)\right\rangle \rightarrow\frac{\hat{L}(\mathbf{Z}_t)\left|\psi(t)\right\rangle }{|| \hat{L}(\mathbf{Z}_t)\left|\psi(t)\right\rangle || },
\end{equation}
where $\hat{L}(\mathbf{Z}_t)$ is the localization operator defined as:
\begin{equation}
\hat{L}(\mathbf{Z}_t)=\frac{1}{(\pi r_{C}^{2})^{3/4}}\,e^{-(\hat{\mathbf{x}}-\mathbf{Z}_t)^{2}/2r_{C}^{2}},\label{eq:L}
\end{equation}
where $\hat{\mathbf{x}}$ is the particle's position operator, $\mathbf{Z}_t$ is a time dependent real valued random variable with probability density
\begin{equation}\label{eq:GRW_pdf}
\mathbb{P}_{\mathbf{Z}_t}(\mathbf{Z}_t=\mathbf{z})=|| \hat{L}(\mathbf{z})\left|\psi(t)\right\rangle || ^{2}
\end{equation}
and $r_{C}$ a second parameter of the model, which determines the width
of the localization gaussian in Eq.~(\ref{eq:L}) and it is usually
taken equal to $r_{C}=10^{-7}$ m. More explicitly, when expressed in the position basis, the collapse described in Eq.~\eqref{collapsegrw} amounts to the change:
\begin{equation}\label{collapsegrw2}
\psi(\mathbf{x}, t)  \rightarrow\frac{1}{\mathcal{N}} \frac{1}{(\pi r_{C}^{2})^{3/4}}\,e^{-(\mathbf{x}-\mathbf{Z}_t)^{2}/2r_{C}^{2}}\psi(\mathbf{x}, t),
\end{equation}
where $\mathcal{N}$ is a normalization factor. 

In general, we have a succession of localization centers $\mathbf{z}_{1}=\mathbf{Z}_{t_{1}}, \mathbf{z}_{2}=\mathbf{Z}_{t_{2}}, \mathbf{z}_{3}=\mathbf{Z}_{t_{3}},...$ selected by $\mathbf{Z}_{t}$ at times $t_{1},t_{2},t_{3},...$, which are generated by the poissonian process $N_t$. We can represent the overall evolution, defined by the two prescriptions (Sch\"odinger equation + collapse events) with a diagram:
\begin{equation}
\left|\psi(t_{init})\right\rangle \overset{Schr\ddot{o}dinger}{\longrightarrow}\left|\psi(t_{1})\right\rangle \overset{collapse}{\longrightarrow}\left|\psi(t_{1}+)\right\rangle \overset{Schr\ddot{o}dinger}{\longrightarrow}\left|\psi(t_{2})\right\rangle \overset{collapse}{\longrightarrow}\left|\psi(t_{2}+)\right\rangle \overset{Schr\ddot{o}dinger}{\longrightarrow}....\label{eq:grw_diagram}
\end{equation}
These two prescriptions can be combined in a single equation, using the formalism of the stochastic analysis~\cite{arnoldbook,hansonbook}:
\begin{equation}\label{eq:grw_wave_function}
d\left|\psi(t)\right\rangle =\left[-\frac{i}{\hbar}\hat{H}dt+\left(\frac{\hat{L}(\mathbf{Z}_t)}{|| \hat{L}(\mathbf{Z}_t)\left|\psi(t)\right\rangle || }-1\right)dN_t\right]\left|\psi(t)\right\rangle. 
\end{equation}
The literature~\cite{PhysRevA.90.062135} usually gives a different prescription for the GRW dynamics in terms of stochastic differential equations. The stochastic differential equation given by Eq.~\eqref{eq:grw_wave_function} is closer to the original idea of GRW and J.S. Bell.

The corresponding statistical operator is defined as:
\begin{equation}
\hat{\rho}(t)=\mathbb{E}_{\mathbf{Z}_t}\left[\mathbb{E}_{\mathbb{Q}}\left[\left|\psi(t)\right\rangle \left\langle \psi(t)\right|\right]\right],
\end{equation}
where the first average $\mathbb{E}_{\mathbb{Q}}$ is taken over the noise $N_t$ while the second average $\mathbb{E}_{\mathbf{Z}}$ over the localization centers $\mathbf{Z}_t$. It is straightforward, using It\^o stochastic calculus, to obtain the equation describing the evolution of the statistical operator $\hat{\rho}(t)$:
\begin{equation}
\frac{d\hat{\rho}(t)}{dt}=-\frac{i}{\hbar}\left[\hat{H},\hat{\rho}(t)\right]+\lambda\left(\intop_{-\infty}^{+\infty}d\mathbf{z}\,\hat{L}(\mathbf{z})\hat{\rho}(t)\hat{L}^{\dagger}(\mathbf{z})-\rho(t)\right),
\end{equation}
which corresponds to the one originally introduced by GRW~\cite{PhysRevD.34.470}.
The generalization to a many-particles system is straightforward: each particle, independently from the others, collapses as described here above.  

In the next section, the GRW dynamics will be derived from Bohmian Mechanics. 

\section{Open quantum system in Bohmian Mechanics and its connection to GRW}\label{sec:The-classical-limit}

We discuss the classical limit of Bohmian Mechanics. The potential problem for describing classical systems in Bohmian Mechanics is that the wave function always obeys the Schr\"odinger equation, therefore it can interfere with itself, also for a macroscopic objects. In such a case, Bohmian Mechanics trajectories typically behave in a highly non-classical way. However, macroscopic objects are always surrounded by some environment, which classicalizes the trajectories \cite{Appleby}. We discuss how this happens. We show how the interaction between a system with an external environment leads to a GRW dynamics for the conditional wave function of an open quantum system, implying that Bohmian trajectories for macroscopic objects follow classical paths, for all practical purposes. 
\begin{figure}[!t]
\begin{center}
\includegraphics[width=1.0\columnwidth,trim={0 0 0 0},clip]{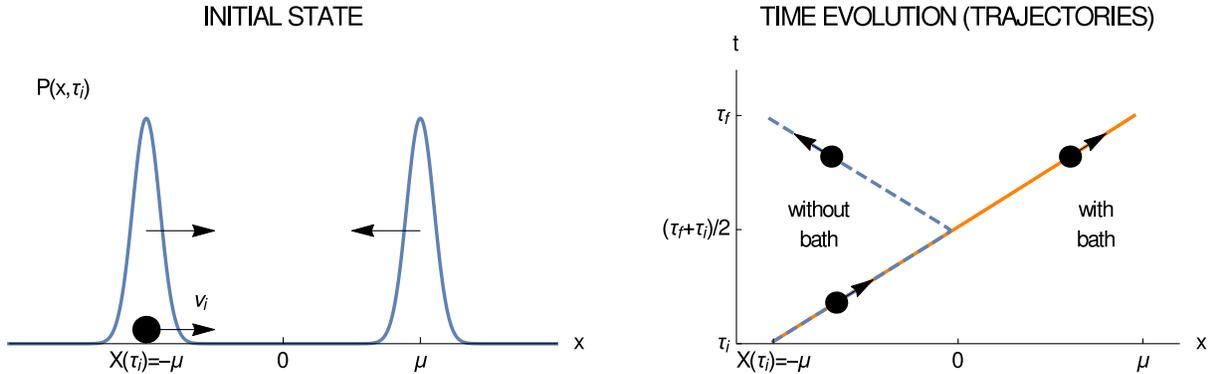}
\caption{Left: We consider a free particle, initially in the superposition $\psi_S(x,\tau_i)$ of two gaussian states moving towards each other, as given by Eq.~\eqref{initial_state}. $P(x,\tau_i)=	\|\psi_S(x,\tau_i)\|^2$  denotes the probability density function of the initial state, while $X(\tau_i)$ denotes the initial particle's position. Right: The evolution of $X(t)$ for an isolated particle (dashed line) and for a particle surrounded by a bath (continuous line). In the former case, the interference between the two wave-packets causes the particle to bounce back. In the latter case, the interaction with the bath immediately collapses the system's conditional wave function near $X(t)$, and effectively classicalizes the trajectory.}
\label{Fig:1}
\end{center}
\end{figure}

To illustrate the potential problem of Bohmian Mechanics in describing the motion of macroscopic objects, and how this is taken care of by decoherence, we consider a macroscopic free particle in one dimension with total mass $M$ described by the initial wavefunction 
\begin{equation}\label{initial_state}
\psi_S(x,\tau_i)=\frac{1}{\sqrt{2}} 
\left(\frac{1}{(2\pi\sigma^2)^{1/4}}e^{-\frac{(x+\mu)^2}{2\sigma^2}+\frac{i}{\hbar}M v_i x} +\frac{1}{(2\pi\sigma^2)^{1/4}}e^{-\frac{(x-\mu)^2}{2\sigma^2} -\frac{i}{\hbar}M v_i x}\right),   
\end{equation}
and by the initial system position $X(\tau_i)$  shown in Fig.~\ref{Fig:1}. As time passes, the two gaussians move towards each other and eventually interfere. In particular, we consider the evolution of such a system, when isolated and when surrounded by a bath of particles. In the former case, when the two gaussians interfere, they guide the position $X(t)$ in a highly non classical trajectory, while in the latter case, the system's conditional wave function, which effectively controls the particle's motion, will localize around $X(t)$ in a short time, eliminating the incoming gaussian from the right, thus classicalizing the trajectory. 

We now begin the derivation. We start with a point particle interacting with only one particle of the environment (which, in the following, we will also refer to as ``bath''). For simplicity, from now on we work in one dimension, but the generalization to three dimensions is straightforward. We consider the total Hamiltonian:
\begin{equation}
\hat{H}=\hat{H}_{\textrm{S}}+\hat{H}_{\textrm{E}}+f_t\hat{V}_{\text{\tiny int}},
\end{equation}
where $\hat{H}_{\textrm{S}}$ and $\hat{H}_{\textrm{E}}$ are the system and environment Hamiltonians, respectively. We assume that the particle of the system and the particle of the bath interact through a von-Neumann type of interaction
\begin{equation}
\hat{V}_{\text{\tiny int}}=\hat{x}_{\textrm{S}}\hat{p}_{\textrm{E}},\label{eq:V}
\end{equation}
where $\hat{x}_{\textrm{S}}$, $\hat{p}_{\textrm{E}}$ are, respectively, the position operator of the system and the momentum operator associated to the particle of the bath. The function $f_t$ identifies the interaction time and it is defined as follows:
\begin{equation}
f_t=\begin{cases}
\frac{1}{t_{f}-t_{i}} \;\; &\text{{if}}\;\;t\in[t_{i},t_{f}],\\
0 &\text{{otherwise}},
\end{cases}
\end{equation}
and is such that, given $g_t:=\int_{t_{i}}^{t}f_s ds$, then $g_{t}=1 \;\forall t  \geq t_f$. 

Let us neglect the evolution of the system and environment given by $\hat{H}_{\textrm{S}}$, $\hat{H}_{\textrm{E}}$, respectively\footnote{Here we make the gross simplification of neglecting the free evolution of both particles. In a more refined approach, which we leave to future research, one can assume that the interaction lasts for a very short time $\Delta t=t_f-t_0$, during which the free dynamics produces a change in the wave function proportional to $\Delta t$. In the limit $\Delta t \rightarrow 0$ (but keeping $g_{t_f}=1$) we have an instantaneous interaction, during which the free dynamics produces no change.}. Then, given the initial total wave function $\psi(x,y,t_i)$, its time evolution is simply given by: 
\begin{equation}
\psi(x,y,t)=\psi(x,y-xg_t,t_i).\label{eq:tot}
\end{equation}
We now make two further assumptions:
\begin{enumerate}
\item The total wave function at the initial time $t_i$ is factorized:
\begin{equation}\label{assumption1}
\psi(x,y,t_i)=\psi_{\textrm{S}}(x,t_i)\psi_{\textrm{E}}(y,t_i),
\end{equation}
where $\psi_{\textrm{S}}$, $\psi_{\textrm{E}}$ are the wave functions of the system and environmental particle, respectively.
\item The initial wave function of the environmental particle is taken equal to:
\begin{equation}\label{assumption2}
\psi_{\textrm{E}}(y,t_i)=\frac{1}{(2\pi\sigma^{2})^{1/4}}e^{-y^{2}/4\sigma^{2}}.
\end{equation}
\end{enumerate}
We now derive the Bohmian velocities. For the system we are considering, Eq.~(\ref{eq:continuity}) takes the form:
\begin{equation}\label{vvv1}
\partial_t|\psi|^{2}=-\partial_x\left(v_{x}|\psi|^{2}\right)-\partial_y\left(v_{y}|\psi|^{2}\right).
\end{equation}
On the other hand, if we write $\partial_t |\psi|^{2}$ starting from the Schr\"odinger equation with $H=f_t\hat{V}_{\text{\tiny int}}$ (i.e. $\hat{H}_{\textrm{S}}=\hat{H}_{\textrm{E}}=0$) we easily get:
\begin{equation}\label{vvv2}
\partial_t|\psi|^{2}=-\partial_{y}\left(f_{t}x|\psi|^{2}\right).
\end{equation}
By comparing Eq.~(\ref{vvv1}) with Eq.~(\ref{vvv2}) we obtain the particles velocities: $v_{x}=0$ and $v_{y}=X(t) f_t$, and thus
by integrating Eq.~(\ref{eq:velocity}) we find the particle trajectories:
\begin{eqnarray}
X(t)&=&X^0, \label{eq:trajectoryX}\\
Y(t)&=&Y^0+X^0g_t \label{eq:trajectoryY},
\end{eqnarray}
where $X^0$, $Y^0$ are, respectively, the initial positions of the system's particle S and that of the environmental particle E. Together, Eqs.~(\ref{eq:tot}), (\ref{eq:trajectoryX}) and (\ref{eq:trajectoryY}) give a complete description of the total system. 

\begin{figure}[!t]
\begin{center}
\includegraphics[width=1.0\columnwidth,trim={0 50 0 80},clip]{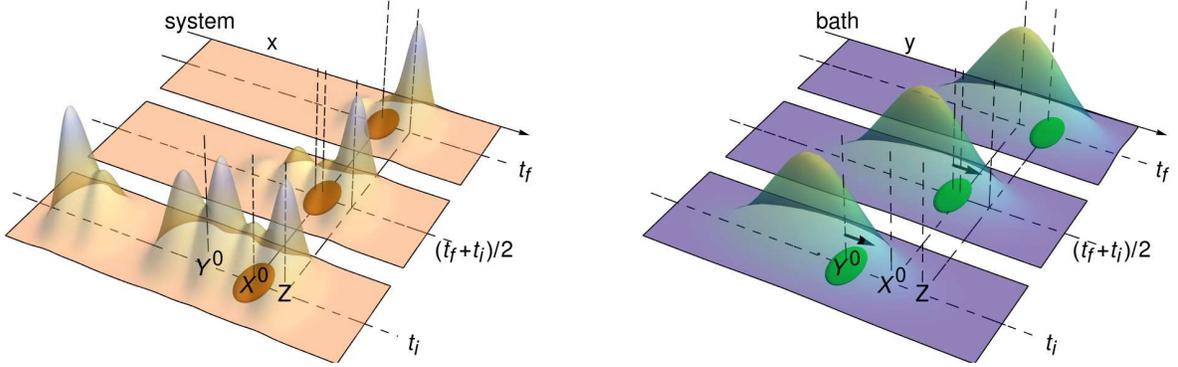}
\caption{Left: time evolution of the position $X(t)$ (orange sphere) and of the conditional wave function $\psi_C(x,t)$ given by Eqs.~\eqref{eq:trajectoryX} and \eqref{psievoultion}, respectively, at times $t_i$, $(t_i+t_f)/2$, $t_f$. The interaction between the system and bath collapses the system's conditional wave function $\psi_C$ near the localization center $Z=X^0+Y^0$. Right: time evolution of the bath particle position $Y(t)$ (green sphere) and of the bath conditional wave function $\psi_B(y,t)$ given by Eqs.~\eqref{eq:trajectoryY} and \eqref{psibathevoultion}, respectively, at times $t_i$, $(t_i+t_f)/2$, $t_f$. The interaction between the system and bath moves the bath's conditional wave function $\psi_B$ near the initial system position $X^0$, while the bath particle position moves from $Y^0$ to $Z=Y^0+X^0$. In both figures, for graphical reasons, the wave functions are normalized such that the maximum values at different times coincide. }
\label{Fig:2}
\end{center}
\end{figure}

Using Eq.~(\ref{eq:conditional}), the conditional wave function for particle S is given by:
\begin{equation}\label{psievoultion}
\psi_{\textrm{C}}(x,t)=\frac{1}{\mathcal{N}}\psi_{\textrm{S}}(x,t_i)\times\frac{1}{(2\pi\sigma^{2})^{1/4}}e^{-(Y(t)-g_t x)^{2}/4\sigma^{2}},
\end{equation}
where $\mathcal{N}$ is a normalization constant. In particular, we see that the wave function at time $t_{f}$, i.e. when the interaction is over, is localized around the point $Z:=Y(t_{f})=X^0+Y^0$,
which we will refer to as the ``localization center''. We can also define the bath particle conditional wave function by setting $x=X(t)=X^0$ in Eq.~\eqref{eq:tot}:
\begin{equation}\label{psibathevoultion}
\psi_{\textrm{B}}(y,t)=\frac{1}{\mathcal{N}'}\psi_{\textrm{S}}(X^0,t_i)\times\frac{1}{(2\pi\sigma^{2})^{1/4}}e^{-(y-g_t X^0)^{2}/4\sigma^{2}},
\end{equation}
where $\mathcal{N}'$ is a normalization constant. The evolution of the positions as given by Eqs.~\eqref{eq:trajectoryX},\eqref{eq:trajectoryY}, as well as the evolutions of the conditional wave functions given by Eqs.~\eqref{psievoultion},\eqref{psibathevoultion} are shown graphically in Fig.~\ref{Fig:2}.

According to the quantum equilibrium hypothesis, the probability density functions of the random variables $X^0$ and $Y^0$ are given, respectively, by $\mathbb{P}(X^0=x)=|\psi_{\textrm{S}}(x,t_i)|^{2}$ and $\mathbb{P}(Y^0=y)=|\psi_{\textrm{E}}(y,t_i)|^{2} =\frac{1}{(2\pi\sigma^{2})^{1/4}}e^{-y^2/4\sigma^{2}}$.
In addition, the variables $X^0$ and $Y^0$ are statistically independent, since the total initial wave function given by Eq.~\eqref{assumption1} is factorized. Hence, according to probability theory,  the probability density function
of the random variable $Z$ is given by the convolution of $\mathbb{P}(X^0=x)$
and $\mathbb{P}(Y^0=y)$~\cite{1751-8121-44-47-478001}:
\begin{equation}\label{eq:bm_pdf}
\mathbb{P}(Z=z)=\intop_{-\infty}^{+\infty}dw\, \mathbb{P}(Y^0=w-z)\mathbb{P}(X^0=w) =\intop_{-\infty}^{+\infty}dw\frac{1}{(2\pi\sigma^{2})^{1/2}}e^{-(w-z)^{2}/4\sigma^{2}}|\psi_{\textrm{S}}(w,t_i)|^{2},
\end{equation}
which corresponds to the GRW prescription in Eq.~(\ref{eq:GRW_pdf}).

In the limit of an instantaneous interaction, the conditional wave function $\psi_{\textrm{C}}(x,t)$ changes instantly according to the GRW prescription given in Eq.~(\ref{collapsegrw2}) with $r_C$ replaced by $\sqrt{2}\sigma$. We give some numerical values, to estimate the magnitude of the collapse. If we take for the environment an ideal gas at thermal equilibrium, a reasonable estimate for $\sigma$ is given by the thermal de Broglie wavelength $\lambda_{th}=\hbar/\sqrt{2\pi m k_{\textrm{B}}T}$, where $m$ is the mass of the gas particle, $k_{\textrm{B}}$ the Boltzmann constant and $T$ the temperature of the gas. For example, if we consider a system interacting with the Earth's atmosphere, which is composed for about $78\%$ by molecular Nitrogen N$_2$, then $\lambda_{th}\simeq 3 \cdot 10^{-12}\textrm{m}$, where we have taken $m=4,7 \cdot 10^{-26}\textrm{kg}$ (the mass of a N$_2$ molecule) and the room temperature $T=298$K. The rate of collisions $\eta$ between the system and the gas can be estimated by using the formula $\eta=n\sigma_{\text{\tiny CS}}\bar{v}$ where $n$ is the density of the gas, $\sigma_{\text{\tiny CS}}$ is the cross section of the interaction between the system and a gas molecule and $\bar{v}$ is the average velocity of the gas molecules \cite{birdbook}. If we consider the interaction of a sphere of radius $R=10^{-3}$m  with a Nitrogen molecule, we have $n=p_{0}/(k_{\textrm{B}}T)=2,46\times10^{25}\textrm{m}^{-3}$ where $p_0=101325\textrm{Pa}$ is the atmospheric pressure at sea level, $\sigma_{\text{\tiny CS}}=\pi R^{2}=3,14\times10^{-6}\textrm{m}^{2}$ (for simplicity we consider the cross-section of classical rigid spheres and we neglect the radius of the gas molecule, being much smaller than the radius $R$ of the system) and $\bar{v}=\sqrt{8k_{\textrm{B}}T/\pi m}=4,72\times10^{2}\textrm{m s}^{-1}$, leading to the rate of collisions $\eta=3,6\times10^{22}\textrm{s}^{-1}$.  The numbers show that the conditional wave function of a (point-like) macroscopic object surrounded by an environment comparable to the Earth's atmosphere is almost immediately localized in space, and remains so in time. Any initial superposition is destroyed. The collapse is equivalent to a GRW collapse with $r_C=10^{-12}\text{m}$ and $\lambda=10^{22}\text{s}^{-1}$.

The generalization to a many-particle environment is straightforward. We assume that the interactions are istantaneous and occur at random times $t_1,\,t_2,\,t_3,...$ governed by a Poisson process $M_t$ with mean value  $\mu$ defined in a probability space $(\Omega,\mathcal{F},\mathbb{P})$, as usually done when discussing collisional decoherence~\cite{openquantumbook}. Let us suppose that at time $t_j$ the system interacts with the $j$-th particle of the bath centered around the point $a_j$ i.e. with wave function (in place of Eq.~\eqref{assumption2}):
\begin{equation}\label{newpsie}
\psi_{\textrm{E}}^j(y,t_j)=\frac{1}{(2\pi\sigma^{2})^{1/4}}e^{-(y-a_j)^{2}/4\sigma^{2}}.
\end{equation}
Then equations for the particle trajectories are given by (in place of Eqs.~\eqref{eq:trajectoryX} and \eqref{eq:trajectoryY}):
\begin{eqnarray}
X(t)&=&X^0, \\
Y_j(t)&=&Y_j^0+X^0g_t=a_j+Y^0+X^0g_t,\label{newy}
\end{eqnarray}
where $Y_j^0$ is distributed as $|\psi_{\textrm{E}}^j|^2$, while $Y^0$ is distributed according to $|\psi_{\textrm{E}}|^2$ with $\psi_{\textrm{E}}$ given by Eq.~\eqref{assumption2}. It is straightforward to verify that the conditional wave function is still given by Eq.~\eqref{psievoultion} (the $a_j$ in Eq.~\eqref{newpsie} cancels with the one in Eq.~\eqref{newy}), thus the localization center is given by the random variable ${Z}_{j}=X(t_{j})+Y^0$. 

To summarize, under the assumptions previously introduced, the conditional wave function of a particle interacting with an external environment undergoes a GRW-type of collapse dynamics as described by Eq.~(\ref{eq:grw_wave_function}). From the previous analysis, it should be clear that the specific form of the ``localization operator" depends on the choice of the interaction Hamiltonian and of the initial wave function of the environment.  
\section{Many particle systems: the classical limit in Bohmian Mechanics}\label{classicallimit2}

Until now we considered only a single point-particle system. To understand the classical limit of Bohmian Mechanics, one has to consider the dynamics of a composite systems. We take a $N$ particle system interacting with the environment. In particular, during the instantaneous interaction between the $k$-th particle of the system with a particle of the environment, the wave function changes to\footnote{The assumption of an instantaneous interaction between a gas particle and the $k$-th particle of the system is fundamental for Eq.~\eqref{change} to be exact. The more realistic case where the interaction holds for a finite time, the $k$-th particle cannot be treated, at least in principle, as free. However, as long as the interaction happens on a time scale which is much shorter than the time scale of the interaction among the $k$-th particle with the other particles of the system which it is linked to, the approximation of instantaneous interaction is good and the dynamics is well described by Eq.~\eqref{change}.}:
\begin{equation}\label{change}
\psi(x_{1},...,x_{N},y,t)=\psi(x_{1},...,x_{N},y-x_{k}\,g_t,t_i).
\end{equation}
We again assume that the system's wave function and bath's wave function at the initial time $t_i$ are factorized:
\begin{equation}\label{assumption1composite}
\psi(x_{1},...,x_{N},y,t_i)=\psi_{\textrm{S}}(x_{1},...,x_{N},t_i)\psi_{\textrm{E}}(y,t_i),
\end{equation}
and that the latter is a gaussian (see Eq.~\eqref{assumption2}).
Similarly to what we did before, from Eq.~(\ref{eq:continuity}) we obtain the particle velocities $(v_{x_{k}}=0$ for $k=1,...,N$, and $v_y=X_k^0f_t$) and thus, by integrating Eq.~(\ref{eq:velocity}), we find the particle trajectories:
\begin{eqnarray}
X_k(t)&=&X_k^0,\label{eq:trajectoriesX}\\
Y_k(t)&=&Y^0+X_k^0\,g_t,\label{eq:trajectoriesY}
\end{eqnarray}
where $X_{1}^0,...,\, X_{N}^0$, $Y^0$ are the initial positions and the label ``$k$'' on $Y_k(t)$ remarks the fact that the environmental particle interacted with the $k$-th particle of the system.

During the interaction the conditional wave function changes to:
\begin{equation}
\psi_{\textrm{C}}(x_{1},...,x_{N},t)=\frac{1}{\mathcal{N}}\psi_{\textrm{S}}(x_{1},...,x_{N},t_i)\frac{1}{(2\pi\sigma^{2})^{1/4}}e^{-(Y_k(t)-g_t x_{k})^{2}/4\sigma^{2}},\label{eq:total_n}
\end{equation}
where $\mathcal{N}$ is again a normalization factor.

We now consider the center of mass wave function for the system.
We introduce the center of mass coordinate $x_{cm}=\frac{x_{1}+...+x_{N}}{N}$
and the relative coordinates $r_{k}=x_{k}-x_{cm}$ with $k=1,...,N-1$. We also construct, from the actual particle positions, the center of mass position 
\begin{equation}\label{com_position}
X_{cm}=\frac{X_{1}+...+X_{N}}{N}
\end{equation}
and the relative particle's positions $R_{k}=X_{k}-X_{cm}$ with $k=1,...,N-1$. We can now define the center of mass wave function conditioned on the relative particle's positions:
\begin{equation}\label{eq:cmwavefunction}
\psi_{cm}(x_{cm},t)=\frac{\psi_{\textrm{S}}(x_{cm},R_{1}(t),...,R_{n-1}(t),t)}{||\psi_{\textrm{S}}(x_{cm},R_{1}(t),...,R_{n-1}(t),t)||}.
\end{equation}
In standard Quantum Mechanics the concept of center of mass wave function can be introduced only when the wave function of the system can be factorized with respect to the center of mass and the relatives coordinates. On the contrary the definition in Eq.~(\ref{eq:cmwavefunction}) is meaningful also for non-factorized wave functions. Moreover, in the case when the wave function can be factorized with respect to the center of mass and the relatives coordinates, the definition in Eq.~(\ref{eq:cmwavefunction}) corresponds to the usual definition of standard Quantum Mechanics. 

We now show that a localization event on the $k$-th particle induces a localization of the center of mass wave function. We note that the exponent in Eq.~(\ref{eq:total_n})
can be written as:
\begin{equation}
Y_k(t)-g_tx_{k}=[Y^0+g_t(X_{cm}^0+R_{k}^0-r_{k})]-g_t x_{cm},
\label{eq:exponential_cm}
\end{equation}
where we label again the values at time $t_i$ with the superscript index $0$.
Hence, from Eqs.~\eqref{eq:cmwavefunction} and \eqref{eq:total_n} and the fact that $R_k(t)=R_k^0$, we obtain the evolution for the center of mass wave function: 
\begin{equation}\label{psicmevolution}
\psi_{cm}(x_{cm},t)=\frac{1}{\mathcal{N}_{cm}}\psi_{cm}(x_{cm},t_i)\frac{1}{(2\pi\sigma^{2})^{1/4}}e^{-(Y_{cm}(t)-g_t x_{cm})^{2}/4\sigma^{2}},
\end{equation}
where $\mathcal{N}_{cm}$ is a normalization factor and we have introduced 
\begin{equation}\label{eq:trajectoryYCM}
Y_{cm}(t):=Y^0+g_tX_{cm}^0. 
\end{equation}
Moreover, it is obvious from Eqs.~\eqref{eq:trajectoriesX} and \eqref{com_position} that
\begin{equation}\label{eq:trajectoryXCM}
X_{cm}(t)=X_{cm}^0.
\end{equation}

As we see, the interaction of a bath particle with a particle of the composite object causes the localization of the center of mass conditional wave function. In particular, Eqs.~\eqref{eq:trajectoryXCM}, \eqref{eq:trajectoryYCM} and \eqref{psicmevolution} have the same form as Eqs.~\eqref{eq:trajectoryX},  \eqref{eq:trajectoryY} and \eqref{psievoultion} for the single particle system. This means the center of mass wave function and center of mass position obey the single particle dynamics with a rescaled $\Lambda$. In general, the rescaling factor depends on the system and how it interacts with the environment. As a first approximation, one can assume that the rate of interactions between the system and an environmental particle increases linearly with the number of particles in the system: $\Lambda=N\lambda$, which is exactly the same amplification mechanism characterizing the GRW model. This is the key feature that allows microscopic particles to behave quantum mechanically and macroscopic objects to behave classically.

In order to complete the discussion of the classical limit within
Bohmian Mechanics, we have to show that the Bohmian trajectory for the center of mass position of a macroscopic system really reduces to a classical trajectory. In particular, we have to show that the center of mass position $X_{cm}(t)$ moves according to Newton's second law. In order to do that, we will refer to another collapse model, the Quantum Mechanics with Universal Position Localizations (QMUPL) model~\cite{PhysRevA.40.1165,PhysRevA.42.5086}. The connection between the GRW and the QMUPL model has been studied in~\cite{trotter}, where it was rigorously proved that the GRW model reduces to the QMUPL model in the limit of infinite rate of localizations ($\lambda \rightarrow \infty$) infinitively weak ($r_C\rightarrow \infty$), with the two limits performed in such a way that the quantity $\lambda_{\text{\tiny{QMUPL}}}:=\lambda/r_C^2$ remains constant. Here we summarize the most relevant results, for our analysis, derived for the QMUPL model in~\cite{doi:10.1142/S0129055X10003886,0305-4470-38-14-008}.

The time evolution of the center of mass wave function of a macroscopic object can be divided into three regimes:
\begin{enumerate}
\item Collapse regime: In a very short time an arbitrary initial wave function localizes in space according to the Born rule. The collapse time, from an arbitrary initial wave function to a wavefunction of spread  $l$, can be quantified as $t_C=\frac{3}{2l^2 \Lambda_{\text{\tiny{QMUPL}}}}$, where $\Lambda_{\text{\tiny{QMUPL}}}=\Lambda/r_C^2$ in an appropriate limit as discussed in~\cite{trotter}. For example, let us consider a sphere of radius $R=1\text{mm}$ in the Earth's atmosphere as described in section~\ref{sec:The-classical-limit}: $(1/\sqrt{2}) r_C = \lambda_{th} = 3\cdot 10^{-12}\text{m}$ and $\Lambda=\eta=3,6\times10^{22}\textrm{s}^{-1}$. We also set $l=10^{-3}\text{m}$. We obtain the estimate  $t_C \approx 10^{-40}\text{s}$. Hence the collapse, for an arbitrary initial wave function, occurs practically immediately.

\item Deterministic regime: For a very long time the wave function can be effectively described by a well localized wave packet with its center moving according to the classical dynamics. The spatial fluctuations around the classical motion, become larger than a given length $L$ in a time $t_{\text{\tiny cl}}= \left(\frac{2}{3}\frac{L}{\sqrt{\Lambda_{\text{\tiny{QMUPL}}}}}\frac{M}{\hbar} \right)^{2/3}$. For example, for the sphere considered here above and $\Lambda=\eta=3,6\times10^{22}\textrm{s}^{-1}$. We also set $M=1\text{g}$ and $L=10^{-3}\text{m}$ and we obtain the estimate $t_{\text{\tiny cl}} \approx 45 \text{min}$.

\item Diffusive regime: Eventually, after a time $t_{\text{\tiny cl}}$ the random motion induced by the Poisson process causes large statistical fluctuations around the deterministic motion, which become more and more significant. 
\end{enumerate}

We are interested in the second regime, which is the only relevant one in all practical physical situations. In such a case, the center-of-mass conditional wave function is extremely well localized: no interference terms appear any longer. More specifically, any initial wave function collapses to an approximately Gaussian state of the form~\cite{0305-4470-38-14-008}: 
\begin{equation}\label{gaussianpsi}
\psi_{cm}(x,t)= \left(\frac{\pi \hbar}{\Lambda M} \right)^{\frac{1}{4}}
\exp \left[ -\frac{z^2}{2}(x-\bar{x}(t))^2+\frac{i}{\hbar} \bar{p}(t) x + i A(t) \right],
\end{equation}
where $z=(1+i) \sqrt{\frac{\Lambda M}{\hbar}}$, while $A(t)$ is a random function of time and $\bar{x}(t)$ and $\bar{p}(t)$ are the mean position and momentum:
\begin{equation}
\bar{x}(t)= \langle \psi(t) | \hat{x}_{cm} | \psi(t) \rangle,\;\;\;\;\;\;\;\bar{p}(t)= \langle \psi(t) | \hat{p}_{cm} | \psi(t) \rangle.
\end{equation}
The Gaussian state in Eq.~\eqref{gaussianpsi} has fixed finite spread both in position and momentum, given by:
\begin{equation}\label{deltaqandp}
\Delta_q =\sqrt{\frac{\hbar}{M \omega}},\;\;\;\;\;\;\;\;\;\;\;\;\Delta_p =\sqrt{\frac{\hbar M \omega}{2}},
\end{equation}
where $\omega=2\sqrt{\frac{\hbar\Lambda}{M}}$. We note that $\Delta_q \Delta_p = \frac{\hbar}{\sqrt{2}}$, which is close to the minimum allowed by the Heisenberg uncertainty relation. 

Let us consider a general center of mass Hamiltonian:
\begin{equation}
\hat{H}^{(s)}_{cm}=\frac{\hat{p}_{cm}^{2}}{2M}+V(\hat{x}_{cm}),\label{eq:sys_hamiltonian}
\end{equation}
where $V(\hat{x}_{cm})$ is the external potential. We assume that over distances $\Delta_q$ the external potential does not vary appreciably and can be approximated as
\begin{equation}
V(\hat{x}_{cm})\simeq V(\bar{x}(t))+\left.\boldsymbol{\nabla} V(x)\right|_{x=\bar{x}(t)}(\hat{x}_{cm}-\bar{x}(t)).
\end{equation}
Hence the Hamiltonian reduces to
\begin{equation}
\hat{H}^{(s)}_{cm} \simeq \frac{\hat{p}_{cm}^{2}}{2M}+\left.\boldsymbol{\nabla} V(x)\right|_{x=\bar{x}(t)}\hat{x}_{cm}\label{eq:sys_hamiltonian2}
\end{equation}
modulo constant terms, which contribute only with a global phase factor. It can be shown, following an analogous calculation as in~\cite{0305-4470-38-14-008, doi:10.1142/S0129055X10003886}, that the mean position $\bar{x}(t)$ and velocity $\bar{v}(t)=\bar{p}(t)/M$ evolve in time as follows:
\begin{align}
\bar{x}(t)&=\bar{x}(t_0)+ \int_{t_0}^{t}\bar{v}(s)ds + \sqrt{\frac{\hbar}{M}}(W(t)-W(t_0)), \label{barxt2}\\
\bar{v}(t)&=\bar{v}(t_0) - \int_{t_0}^t \frac{\left.\boldsymbol{\nabla} V(x)\right|_{x=\bar{x}(s)}}{M} ds+ \sqrt{\Lambda}\frac{\hbar}{M}(W(t)-W(t_0)), \label{barvt2}
\end{align}
where $\bar{x}(t_0)$, $\bar{v}(t_0)=\bar{p}(t_0)/M$ are the initial values and $W(t)$ is a standard Wiener process. The fluctuations can be estimated by setting $W(t)\sim \sqrt{t}$. In particular, for a sphere of radius $R=1\text{mm}$ and mass $M=1\text{g}$ moving in Earth's atmosphere as described before, the position fluctuations are
\begin{align}
\sqrt{\Lambda}\frac{\hbar}{M}\int_{t_0}^{t} W(s) ds &\simeq \frac{2}{3} \sqrt{\Lambda}\frac{\hbar}{M} (t^{3/2}-t_0^{3/2}) \simeq \left[ 10^{-20} \text{m}\text{s}^{-3/2} \right] (t^{3/2}-t_0^{3/2})\\
\sqrt{\frac{\hbar}{M}}(W(t)-W(t_0)) &\simeq \sqrt{\frac{\hbar (t-t_0)}{M}} \simeq \left[ 10^{-16} \text{m}\text{s}^{-1/2} \right] (t-t_0)^{1/2} , 
\end{align}
while the velocity fluctuations are:
\begin{equation}
\sqrt{\Lambda}\frac{\hbar}{M}(W(t)-W(t_0)) \simeq \frac{\hbar}{M}\sqrt{\Lambda (t-t_0)} \simeq \left[ 10^{-20} \text{m}\text{s}^{-3/2} \right] (t-t_0)^{1/2}.
\end{equation}
Neglecting these fluctuations, which are small for macroscopic objects, we obtain the  deterministic equations:
\begin{align}
\bar{x}(t)&=\bar{x}(t_0)+ \int_{t_0}^{t}\bar{v}(s)ds, \label{barxt22}\\
\bar{v}(t)&=\bar{v}(t_0) - \int_{t_0}^t \frac{\left.\boldsymbol{\nabla} V(x)\right|_{x=\bar{x}(s)}}{M} ds, \label{barvt22}
\end{align}

We can now extract the Bohmian velocity from Eq.~(\ref{eq:continuity}) using the asymptotic wave function $\psi_{cm}(x,t)$ given by Eq.~\eqref{gaussianpsi}:
\begin{equation}
\partial_t\left| \psi_{cm}(x,t)\right|^2=
-\partial_x
\left( \left|\psi_{cm}(x,t)\right|^2 \frac{d\bar{x}(t)}{dt} \right),
\end{equation}
hence the Bohmian velocity for the center of mass position $X_{cm}(t)$ is given by:
\begin{equation}\label{bm_v_classical}
\frac{dX_{cm}(t)}{dt}=\frac{d\bar{x}(t)}{dt}=\bar{v}(t),
\end{equation}
with $\bar{v}(t)$ given by Eq.~\eqref{barvt22}. In particular, we notice that the velocity $\frac{dX_{cm}(t)}{dt}$ for the center of mass position coincides with mean velocity $\bar{v}(t)$ of the asymptotic gaussian given by Eq.~\eqref{gaussianpsi}, regardless of the initial position $X_{cm}^0$ (when stochastic fluctuations are neglected as assumed here)\footnote{This can be explained by noting that the asymptotic gaussian has finite spread (see Eq.~\eqref{deltaqandp}) and that two different realizations of Bohmian trajectories  cannot cross. This, known as the "no-crossing theorem" in Bohmian Mechanics, is a direct consequence of the equation of motion for particle positions given by Eq.~\eqref{velvx}. In particular, since this differential equation is first order in time, the velocity field $v(x,t)$ has a unique value for a given value of $t$ and $x$, thus not allowing the crossing of trajectories.)}.
 
In addition, we also assume $X(t) \simeq \bar{x}(t)$, since the spread in position given by Eq.~\eqref{deltaqandp} is very small for a macroscopic object. As an example, for the sphere of radius $R=1\text{mm}$ and mass $M=1\text{g}$ moving in the Earth's atmosphere, we have $\Delta_q \simeq 10^{-14} \text{m}$, which means that the wave function is a Dirac delta for all practical purposes. Hence from Eqs.~\eqref{barvt22} and \eqref{bm_v_classical} we obtain:
\begin{equation}
\frac{dX_{cm}^2(t)}{dt^2}= -\frac{1}{M} \left.\boldsymbol{\nabla} V(x)\right|_{x=X_{cm}(t)},
\end{equation}
which is Newton's second law.

\section{Discussions and conclusions \label{sec:Collapse-models-revisited}}

Our analysis shows how Bohmian trajectories become classical when particles glue together to form macroscopic objects, which unavoidably interact with the surrounding  environment. Interestingly enough, it shows that the dynamics of the conditional wave function obeys to a collapse type of equation which, under reasonable assumptions, is that of the GRW model. In addition, for the center of mass conditional wave function of a composite system, an amplification mechanism is present, exactly like in the GRW model. 

Our analysis is interesting also in the context of collapse models. In our case the instantaneous wave function collapse postulated by GRW is the coarse-grained version of a continuous interaction between the system and an external agent. This might open the way to finding an underlying theory, out of which GRW (and collapse models in general) emerges in a suitable coarse-grained manner, as first suggested in~\cite{tracebook}. This might also resolve the issue of the lack of energy conservation in collapse models. According to our scheme, it would not be a fundamental property of the dynamics: the global energy, that of the system and the external agent causing the collapse, would be preserved.      

\begin{acknowledgments}
The authors acknowledge financial support from FRA 2013 (UNITS). The authors thank Prof. Detlef D\"{u}rr, Dr. Ward Struyve and Davide Romano for many useful discussions.
\end{acknowledgments}

\bibliography{bm_grw}

\end{document}